\documentclass{emulateapj}

\usepackage{natbib}

\slugcomment{Accepted for publication in  {\it The Astrophysical Journal}}

\shorttitle{Reverse Shock Evolution in  \sn1006}
\shortauthors{Winkler, et al.}

\begin{document}


\newcommand{\vdag}{(v)^\dagger}


\newcommand\lam{\mbox{$\:\lambda $ }}
\newcommand\lamlam{\mbox{$\:\lambda\lambda $ }}
\newcommand\ha{{H$\alpha$}}
\newcommand\hb{{H$\beta$}}

\newcommand\BT{B_T}
\newcommand\MBT{M_{B_T}}
\newcommand\Mstar{M^*}
\newcommand\Lstar{L^*}
\newcommand\Mpc{\:\rm{Mpc}}
\newcommand\kpc{\:\rm{kpc}}
\newcommand\pc{\:\rm{pc}}
\newcommand\kms{\:\rm{\,km\,s^{-1}}}
\newcommand\kmsy{\:\rm{\,km\,s^{-1}\; yr^{-1}}}
\newcommand\asy{\:\rm{\,arcsec\,yr^{-1}}}
\newcommand\masy{\:\rm{\,mas\:yr^{-1}}}
\newcommand\rfid{r_{\rm{fid}}}

\newcommand\arcspix{\:\arcsec\:{\rm pixel}^{-1}}
\newcommand\perpix{\:{\rm pixel}^{-1}}
\newcommand\LUM{\:{\rm ergs\:s^{-1}}}
\newcommand\FLUX{\:{\rm ergs\:cm^{-2}\:s^{-1}}}
\newcommand\FLUXARCSEC{\:{\rm ergs\:cm^{-2}\:s^{-1}\:arcsec^{-2}}}
\newcommand\FLUXARCMIN{\:{\rm ergs\:cm^{-2}\:s^{-1}\:arcmin^{-2}}}
\newcommand\FLUXSR{\:{\rm ergs\:cm^{-2}\:s^{-1}\:sr^{-1}}}
\newcommand\COUNTS{\:{\rm counts\:s^{-1}}}
\newcommand\CPM{\:{\rm counts\:s^{-1}\:arcmin^{-2}}}
\newcommand\VEL{\:{\rm km\:s^{-1}}}
\newcommand\ETAL{{\it et\:al.}}
\newcommand\OIGS{\:{\rm ergs\:cm^{-2}\:s^{-1}\:\AA^{-1}}}
\newcommand\LA{Lyman\thinspace$\alpha$}

\newcommand\eg{{\it e.g.}}
\newcommand\ie{{\it i.e.}}
\newcommand\etal{et\thinspace al.}
\newcommand\ergflux{ergs\thinspace cm$^{-2}$\thinspace s$^{-1}$\ }
\newcommand\sn{{SN\thinspace 1006}}


\newcommand\hii{\ion{H}{2}}
\newcommand\sii{[\ion{S}{2}]}
\newcommand\oiii{[\ion{O}{3}]}
\newcommand\feii{\ion{Fe}{2}}
\newcommand\feiii{\ion{Fe}{3}}
\newcommand\siii{\ion{S}{3}}
\newcommand{\SIii}{\ion{Si}{2}}
\newcommand{\SIiii}{\ion{Si}{3}}
\newcommand{\SIiv}{\ion{Si}{4}}
\newcommand\caii{\ion{Ca}{2}}

\newcommand{\sicosfig}{
    \begin{figure}[t!]
    \begin{center}
    \leavevmode
    \includegraphics[scale=.6]{si1260_cos}
    \end{center}
    \caption[1]{
    \label{sicos}
COS absorption spectrum of SN~1006
around the redshifted \SIii\ $1260 \, {\rm\AA}$ feature,
showing the best fit Gaussian profile of shocked \SIii,
and the residual unshocked \SIii.
The spectrum is the ratio of
the interstellar-line-excised,
continuum-corrected
COS spectrum of the SM star,
and the STIS spectrum of the comparision star PG\ 0839+399,
each smoothed to a resolution of $80 \kms$ FWHM
before their ratio was taken.
The upper spectrum shows the
interstellar-line-excised,
continuum-corrected comparison stellar spectrum
of PG\ 0839+399 at the same resolution,
shifted by
$+ 32 \kms$ to mesh its stellar lines with those of the SM star,
and offset to separate it from the SN~1006 spectrum.
    }
    \end{figure}
}


\title{Time Evolution of the Reverse Shock in SN 1006\footnote{This paper is based on observations from the NASA/ESA {\it Hubble Space Telescope}, obtained at the Space Telescope Science Institute, which is operated by AURA, Inc., under NASA contract NAS5-26555.  These observations are associated with program 11659.}}


\author{P. Frank Winkler}
\affil{Department of Physics, Middlebury College, Middlebury, VT 05753}
\email{winkler@middlebury.edu}

\author{Andrew J. S. Hamilton}
\affil{JILA and Dept. of Astrophysical \& Planetary Sciences, University of Colorado, Boulder CO 80309}
\email{andrew.hamilton@colorado.edu}

\author{Knox S. Long}
\affil{Space Telescope Science Institute, Baltimore MD 21218}
 \email{long@stsci.edu}

\author{Robert A. Fesen}
\affil{Dartmouth College, Hanover NH 03755}
\email{Robert.Fesen@snr.dartmouth.edu}


\begin{abstract}

The Schweizer-Middleditch star, located behind the SN~1006 remnant and near its center in projection, provides the opportunity to study cold, expanding ejecta within the SN~1006 shell through UV absorption.  Especially notable is an extremely sharp red edge to the \SIii\ 1260 \AA\ feature, which stems from the fastest moving ejecta on the far side of the \sn\ shell---material that is just encountering the reverse shock.   Comparing {\it HST} far-UV spectra obtained with COS in 2010 and with STIS in 1999, we have measured the change in this feature over the intervening 10.5-year baseline.  We find that the sharp red edge of the \SIii\ feature has shifted blueward by $0.19\pm 0.05$ \AA, which means that the material hitting the  reverse shock in 2010 was moving slower by $44 \pm 11 \kms$ than  the  material  that was hitting it in 1999, a change corresponding to $- 4.2 \pm 1.0 \kmsy$.  This is the first observational confirmation of a long-predicted dynamic effect for a reverse shock: that the shock will  work its way inward through expanding supernova ejecta and encounter ever slower material as it proceeds.  We also find  that the column density of {\it shocked} \SIii\ (material that has passed through the reverse shock) has decreased by $7 \pm 2\%$ over the ten-year period.   The decrease could indicate that in this direction the reverse shock has been ploughing through a dense clump of Si, leading to pressure and density transients.

\end{abstract}




\section{Introduction}

The idea of a reverse shock that can reheat expanding ejecta in a young supernova remnant (SNR) was  introduced in a pair of companion papers by \citet{gull73a} and  \citet{rosenberg73}, and was further articulated by \citet{mckee74} and by \citet{gull75}.   It was  rapidly incorporated into the interpretation of early X-ray data from young SNRs \citep[\eg,][]{charles75}, and has become firmly entrenched into  the canonical model of a young SNR\@.    Briefly, the idea is that following a supernova (SN) event, a blast wave moves outward through the surrounding interstellar or circumstellar material, which becomes swept up behind the expanding strong shock.   Within this shell, a contact discontinuity forms between the expanding SN ejecta and the newly swept-up material, and an internal shock develops that works its way inward  through the ejecta, reheating them and giving rise to strong X-ray and optical emission from material that is highly enriched in heavy elements.  As the SNR evolves and the reverse shock works its way inward, it should encounter ever-slower ejecta---a proposition that we test through the observations reported here.

Despite this long pedigree, detailed observations of the development of the reverse shock over time in actual young SNRs have been sparse.  The most detailed has been for Cas~A:  the {\it Chandra} images clearly showing the reverse shock and its encounter with ejecta of different compositions \citep {hwang00, hwang04} and the {\it Hubble} Space Telescope ({\it HST}) images showing the evolution of individual ejecta knots  as they encounter the reverse shock \citep{morse04, fesen11}.   In this paper, we report the results of an experiment designed to test a definite prediction of the reverse-shock model: that as the reverse shock moves inward through the freely-expanding ejecta, the velocity of material entering the reverse shock gradually decreases over time.   The freely expanding ejecta are too cold to be readily detected in emission, but the velocity profile can be probed  through absorption spectroscopy carried out in the UV, where ions commonly found in ejecta have strong resonance lines connecting to the ground state.   One of very few objects where this technique has so far been employed is the remnant of the bright Type Ia supernova SN~1006 (SNR G327.6+14.6), located    at a distance of 2.2 kpc and   $\sim 500$ pc  above the Galactic plane \citep{winkler03}.   Located only 2\farcm8 arcmin south of the projected center of the the 15\arcmin\ radius SN~1006 shell, the UV-bright  \citet[][henceforth SM]{schweizer80} star enables  study of a ``core sample" through the remnant.\footnote{In addition to the SM star, \citet{winkler05} have used two fainter background QSOs to probe additional lines of sight through SN~1006.}

The  SM star is an OB subdwarf,  probably not much farther away than SN~1006.  It is relatively bright ($V = 16.74,\ B-V = -0.14$), with low foreground extinction, $E(B-V)=0.1$ \citep{schweizer80, wu93}.   Its potential as a background ``UV lightbulb" for probing the interior of SN~1006 through absorption spectroscopy was first exploited by \citet{wu83} using the International Ultraviolet Explorer ({\it IUE}), with subsequent studies carried out by \citet{fesen88, wu93, wu97}, and \citet{blair96}.   These spectra show strong \feii\ lines  at 2383 \AA\ and 2600 \AA\ that are broad and nearly symmetric, and in addition  lines from  \SIii, \SIiii, and \SIiv, all of which are also  broad, but with profiles that are entirely at positive (redshifted) velocities \citep{wu97}.    The most detailed UV spectrum of the SM star was obtained from the {\it Hubble} Space Telescope ({\it HST}) in 1999 with the Space Telescope Imaging Spectrograph (STIS)  and reported by \citet[][hereafter HFB07]{hamilton07}, which also summarizes the previous UV studies.

The strongest feature in the 1150--1700 \AA\ band is broad absorption due to \SIii\ 1260.4 \AA, all on the red side of the line, extending from a velocity of  $\sim+ 2400 \kms$ to a sharp edge at $+7026\kms$ (HFB07).   \citet{hamilton97, hamilton07}  have modeled this profile as a combination of both shocked and unshocked \SIii\ ejecta within SN~1006: the highest redshifts result from ejecta that are still freely expanding on the far side of the shell, with the sharp edge due to material just before it enters the reverse shock.   The reverse-shock model predicts that, over time, the shock encounters ever slower-moving ejecta, and hence the red edge should gradually move to shorter wavelength.  We report here a second-epoch spectrum of the SM star obtained with the Cosmic Origins Spectrograph (COS) on {\it HST} that, when compared with the 1999 STIS spectrum,  demonstrates this phenomenon for the first time.

\section{Observations and Wavelength Registration} 

For this study, we used as the first-epoch spectrum the far-ultraviolet echelle spectrum of the SM star  obtained with STIS in July 1999 and described by HFB07.  Briefly, the STIS observation used a $2\arcsec \times 2\arcsec$ slit, the E140M grating and the FUV-MAMA detector to cover wavelengths 1150--1700 \AA\ at a scale of  0.0122 \AA $\perpix$.   The line spread function is non-Gaussian, with a narrow core and broad wings, and in the crucial region of the spectrum near 1300 \AA\ has a FWHM of $4.6\kms$.   

For the second-epoch observation, we used the newly installed COS instrument, since it has a far higher throughput than STIS, though  with spectral resolution that is almost four times lower.   The observation was carried out on 2010 Jan 16 using the  $2\farcs 5$ diameter primary science aperture and the G130M grating.  A central wavelength setting of 1327 \AA\, and standard subsampling (FP-POS = auto) were used to cover the range $\sim$ 1170--1470 \AA, with a  gap extending over 1317--1327 \AA, at a dispersion of 0.010 \AA$\;\perpix$.   The COS line spread function is also non-Gaussian, with FWHM 18$\kms$  near 1300 \AA\ \citep[see][for a summary of the COS on-orbit performance]{sahnow10}.    Because of the higher throughput with COS, only a single orbit was required to obtain a spectrum with signal-to-noise similar to that obtained in eight orbits with STIS\@.  Key aspects of the two observations are summarized in Table 1, and   
in Fig.\ 1 we plot the spectra from both observations, binned to emphasize the broad features.    The time separation between the epochs is 10.48 yr.

\notetoeditor{We suggest that Fig. 1 be a full-page width (2 column) figure about here.}

In order to measure the small shift in the wavelength of the absorption edge, it is crucial that the STIS and COS spectra be precisely registered in wavelength.  In order to achieve more precise registration than can be done using the wavelength scale provided with pipeline-processed COS data, we can use the many narrow interstellar and stellar lines in the wavelength region near the edge.   
We have used a cross-correlation analysis (the IRAF\footnote{IRAF is distributed by the National Optical Astronomy 
Observatories, which is operated by the  AURA, Inc. under cooperative 
agreement with the National Science Foundation.}  task {\tt fxcor}), first to cross-correlate  each of the four individual (subsampled) frames, made tiny wavelength shifts  to best align these, and averaged them to optimize the total COS spectrum.  We then used a similar cross-correlation analysis  to determine the small offsets between the COS and STIS spectra on both the blue (1245 - 1264 \AA) and red (1292 - 1301.5 \AA) sides of the edge, and found that with the original wavelength scales, the narrow lines differ in wavelength by as much as $40$ m\AA, or $ 9.6\kms $.  The differences are  systematic, so we were able to   apply a simple linear transformation to the COS wavelength scale to achieve registration with the STIS one to within $\pm 6$ m\AA, or $\pm 1.5 \kms $\@.   In Fig.\ 2 we plot the difference between measured  wavelengths in the COS and STIS spectra for several narrow lines, both before and after the transforming the COS spectrum.
 After registering the COS and STIS spectra, we find a clear shift in the red edge of the 1260 \AA\ absorption feature to shorter wavelength, which we have measured as described in the next section.



\section{Model Fit to the COS Spectrum}
\label{model}

As discussed by
\citet{hamilton97, hamilton07},
the unusual shape of the broad, redshifted \SIii\ $1260 \, {\rm \AA}$
absorption feature in SN~1006
can be understood as arising from a combination of unshocked and shocked \SIii\ ejecta.
The sharp red edge at $\sim 7000 \kms$ marks the position of a reverse shock front,
where unshocked \SIii\ is ``instantaneously'' decelerated and thermalized.
The lifetime of the resulting shocked \SIii\ against collisional ionization
is of the order of the age of the remnant,
so it is not surprising that shocked \SIii\ should survive to remain observable.
While separated three-dimensionally of course, both shocked and unshocked \SIii\ lie along the line of sight.

Fig.~\ref{sicos}
shows the result of fitting the $1260 \, {\rm \AA}$ feature observed with COS
to a combination of unshocked and shocked \SIii\@. The analysis is essentially the same as that of HFB07,
to which the reader is referred for more detail.
Fig.~\ref{sicos}
can be compared to Figure\ 4 of HFB07,
which showed a similar fit to the same feature observed with STIS\@.

In brief,
each of the observed spectra were interpolated across narrow
interstellar lines, and their continua fitted to a low-order polynomial.
In the case of the STIS spectra of the SM star and the comparison
star PG\ 0839+399 analysed in 2007,
the continuum was a quintic polynomial in log flux versus inverse wavelength
fitted over unabsorbed regions from $1150$ to $1700 \, {\rm \AA}$.
The COS and STIS spectra are in generally excellent agreement
(see \S\ref{absorption}),
but there is a small but significant difference in the overall
level of fluxes.
To compensate for this difference,
we multiplied the COS spectrum by a factor linear
in log flux versus inverse wavelength,
varying from
$1.03$ at $1190 \, {\rm \AA}$
to
$1.08$ at $1290 \, {\rm \AA}$.

Fig.~\ref{sicos} also
shows the ratio of the interstellar-line-excised, continuum-corrected
spectrum of the SM star observed with COS,
to that of the comparison star PG\ 0839+399 observed with STIS.
The comparison star PG\ 0839+399
\citep*{green86}
was identified by
\citet{blair96}
as having not only a similar temperature and gravity,
but also similar photospheric abundances 
and low extinction.
As remarked by HFB07,
the spectrum of the comparison star is not as similar in detail
to that of the SM star as one might have liked,
but dividing the SM spectrum by the comparison one 
does help to clarify which features in the SM star spectrum are due to SN~1006.


Three separate velocity parameters can be measured from the line profile
shown in
Fig.~\ref{sicos}:
the free-expansion velocity $r_s/t$
(shock radius $r_s$ divided by age $t$)
of unshocked \SIii\ at the reverse shock front, and 
 the mean velocity $ \overline v$
and  one-dimensional velocity dispersion $\sigma$ of shocked \SIii.
Not surprisingly,
in view of the good agreement between the STIS and COS spectra,
and the similar analysis,
the parameters measured from COS are quite similar to those
measured from STIS,
Table~\ref{par}.

The three velocities are related by shock jump conditions.
As noted by HFB07,
the measured velocities point to
essentially all of the shock energy going into heating ions,
with little energy in heating electrons,
consistent with other observational evidence
\citep{laming96,ghavamian02,vink03,vink05},
and also with little energy in accelerating particles
\citep[e.g.][]{kosenko11}.
In the absence of electron heating or particle acceleration,
the change $\Delta v $ 
in the velocity between the unshocked and shocked \SIii\
should equal the three-dimensional velocity dispersion $\sqrt{3} \sigma$
of the shocked \SIii,
\begin{equation}
\label{Deltav}
  \Delta v \equiv \frac{{r_s }}{t}  - \bar{v} = \sqrt{3} \, \sigma
  \ .
\end{equation}
If some of the shock energy went into electron heating or particle acceleration,
then the velocity dispersion would be correspondingly smaller.
If all three velocities are treated as free parameters,
the observations mildly prefer the opposite,
an ion velocity dispersion slightly ($1 \sigma$) larger than predicted
by equation~(\ref{Deltav}).
In the present paper
we choose to constrain the fit,
Fig.~\ref{sicos}, 
such that equation~(\ref{Deltav}) is satisfied,
whereas HFB07
chose to measure all three velocities separately from the STIS spectrum.
HFB07
estimated a $3\sigma$ upper limit of $0.26$
of the shock energy going into forms other than ion heating.

Fits constrained to satisfy equation~(\ref{Deltav})
prefer slightly less shocked Si~II absorption at its peak near $5{,}000 \kms$.
A simple way to accomodate the preference
is to fit the blue edge of the profile up to $5{,}100 \kms$
instead of the $5{,}600 \kms$ adopted by HFB07.
The values quoted in Table~\ref{par} are best fits over the reduced range.
The quoted uncertainties
are $1 \sigma$ statistical uncertainties about the best fit,
subject to variations 
{\em not\/} constrained by equation~(\ref{Deltav}).

The observed column density of shocked \SIii\ measured from the line
profile in Fig.~\ref{sicos} is
\citep[compare eq.~8 of][]{hamilton97}
{\small
\begin{equation}
\label{NshkSiIIobs}
  N^{\rm shk}_{\rm Si II}
  =
  {m_e c \over \pi e^2 f \lambda} \sqrt{2\pi} \, \sigma \tau_0\\
  =
  7.4 \pm 0.3 \times 10^{14}
  \, {\rm cm}^{-2}
  \  (\mbox{measured}),
\end{equation}
}
where
$\lambda = 1260 \, {\rm \AA}$ is the wavelength of the line,
$f = 1.007$
\citep{morton1991}
is its oscillator strength,
$\tau_0$ is the optical depth at line center
of the fitted Gaussian profile of shocked Si~II,
and the factor $\sqrt{2\pi}$ comes from integrating over the profile.
The uncertainty in the estimate~(\ref{NshkSiIIobs})
is the statistical uncertainty arising from photon shot noise,
and does not include ``systematic'' uncertainty arising from
modeling choices, which include placement of the continuum,
and the range of velocities over which the fitting is done.
The systematic uncertainty is somewhat larger.
Notably, if we extend the range
over which we fit the blue edge of the shocked \SIii\ profile 
to $5{,}600 \kms$ (as in HFB07)
instead of $5{,}100 \kms$,
then the measured column density increases to
$8.0 \times 10^{14} \, {\rm cm}^{-2}$.
The measured value~(\ref{NshkSiIIobs})
is listed in Table~\ref{par},
along with the value that
HFB07
would have measured from STIS
but did not report.

The measured column density~(\ref{NshkSiIIobs}) of shocked Si~II
is consistent with that predicted by steady ionization of Si~II
\citep[compare eq.~13 of][]{hamilton97},
{\small
\begin{eqnarray}
\label{NshkSiII}
  N^{\rm shk}_{\rm Si II} &  = &  {n_{\rm SiII} v_s \over 4 n_e \langle \sigma v \rangle_{\rm Si II}} \nonumber \\
  				     &  = &  6 \pm 2 \times 10^{14}\, {\rm cm}^{-2}\left({n_e / n_{\rm Si II} \over 2.5}\right)^{-1}(\mbox{predicted}),
\end{eqnarray}
}
where $\langle \sigma v \rangle_{\rm Si II}$
is the collisional ionization rate (see below),
and the quoted uncertainty is a nominal estimate of the
uncertainty in that rate.
The electron to Si~II ratio of
$n_e / n_{\rm Si II} = 1.5 / 0.6 = 2.5$
in equation~(\ref{NshkSiII})
is a fiducial value that assumes
that each Si ion contributes on average
1.5 electrons during its steady state ionization,
and that the Si is in a silicon-rich region
where it comprises 60\% of the ion abundance by number,
the remaining 40\% being mostly S, consistent
with SN\,Ia nucleosynthesis models
\citep{hoflich1998}.\footnote{Many lines of evidence suggest that SN~1006 was a Type Ia event \citep[e.g.,][]{winkler03}.}

As discussed by \citet{hamilton97},
it is easy to reduce the theoretical
value~(\ref{NshkSiII}) of shocked \SIii,
by diluting the Si with other elements
(i.e.\ by increasing $n_e / n_{\rm Si II}$),
or by truncating the Si downstream,
but it is harder to increase the theoretical value,
to approach the slightly greater
observed value~(\ref{NshkSiIIobs}).
Thus agreement between the measured and predicted column densities of shocked Si~II
appears to require that the absorbing shocked Si arises from silicon-rich ejecta.
This conclusion is consistent with the absence of Fe~II absorption
with the same profile as the shocked \SIii.
Although \feii\ absorption is observed,
its profile is dramatically different from that of the shocked \SIii,
indicating that little iron is mixed in with the shocked silicon
\citep{hamilton97}.

In \citet{hamilton97}
there seemed to be a possible discrepancy between the observed
and predicted values of the column density of shocked Si~II.
That discrepancy has now disappeared,
in part thanks to a reduction in the collisional ionization cross-section.
The collisional ionization rate used in equation~(\ref{NshkSiII}) is
$\langle \sigma v \rangle_{\rm Si II} = 4.4 \times 10^{-8} \, {\rm cm}^3 \, {\rm s}^{-1}$,
obtained by integrating cross-sections
from \citet{clark2003}
over a Maxwellian distribution of colliding electrons
at temperature $95 \, {\rm eV}$,
the temperature reached by electrons as a result
of Coulomb collisions with Si~II ions
over the collisional ionization timescale of Si~II.
The collisional ionization rate from \citet{clark2003}
is smaller than the rate
$6.1 \times 10^{-8} \, {\rm cm}^3 \, {\rm s}^{-1}$
from \citet{lennon1988} used by \citet{hamilton97}.
The newer rate, based on theoretical computations,
should be more reliable than the older rate,
based on isoelectronic scaling.

A concern is that the collisional ionization time $t_{\rm Si II}$ of Si~II
is comparable to the age $t$ of the remnant
(specifically,
$t_{\rm Si II} / t \approx 0.8$,
given the pre-shock Si~II density from
the observed optical depth of freely expanding unshocked Si~II
at the shock front,
an estimated electron-to-Si~II density
$n_e / n_{\rm Si II} \approx 2.5$,
and the theoretical collisional ionization rate
\citep[cf.][]{hamilton97}).
The collisional ionization time is long enough that
the shocked Si~II absorption
should be a superposition of components
shocked at various times with various velocities.
It is somewhat surprising that the profile should fit well
to a single component satisfying the shock jump conditions.
A possible explanation of the coincidence is that
one-dimensional hydrodynamic models predict rather flat
velocity and temperature profiles of shocked ejecta
\cite[e.g.][]{dwarkadas98,kosenko11}.
Another possible explanation,
suggested by the faster than expected
change in the optical depth of shocked Si~II,
\S\ref{absorption},
is that the absorption is being produced by a lump of silicon-rich ejecta
whose average density was higher than that observed today,
and which has been shocked over a timescale shorter than the age of the remnant.


\section{Evolution of the Shock Front from STIS to COS}
\label{evolution}

As the reverse shock propagates into the \SIii\ ejecta,
the free-expansion velocity $r_s/t$
of unshocked gas entering the shock at radius $r_s$
should decrease with time $t$ as 
\begin{equation}
\label{velocityevolution}
  \frac{{d\left( {r_s /t} \right)}}{{dt}} =  - \frac{v_s}{t} \ , 
\end{equation}
where $v_s$ is the velocity of the reverse shock relative to the ejecta.
The shock velocity $v_s$ is related to the change $\Delta v$
in the velocity between unshocked and shocked \SIii\ by
$v_s = (4 / 3) \Delta v$
(regardless of electron heating).
The deceleration $\Delta v$
measured from the COS spectrum shown in Fig.~\ref{sicos}
yields a shock velocity of
$v_s = 2630 \pm 120 \kms$,
in good agreement with the STIS measurement of
$v_s = 2680 \pm 120 \kms$.
As described by HFB07,
equation~(\ref{velocityevolution})
coupled with the measured shock velocity
predicts that the free-expansion velocity at the shock
should be decreasing at the rate of
$- 2.7 \pm 0.1 \kmsy$.
The predicted change
$\Delta ( r_s / t )$
in the free-expansion velocity of the shock
over the 10.5 year interval between the STIS and COS observations
is then
\begin{equation}
\label{predictedDelta}
  \Delta ( r_s / t )
  = - 28 \pm 1 \kms 
  \quad
  (\mbox{predicted}).
\end{equation}

Fig.~\ref{cosstisfit}
compares the COS and STIS spectra
in the vicinity of the putative reverse shock front
near $7000 \kms$.
To permit direct comparison of the spectra,
the COS spectrum has been convolved with the STIS
line spread function (LSF),
while the STIS spectrum has been convolved with the COS LSF.
The net resolution is essentially that of COS,
since the COS resolution ($18 \kms$ FWHM)
is over three times broader than STIS ($4.6 \kms$ FWHM).
The flux in the COS spectrum has been multiplied by
a slowly-varying factor, about equal to
$1.08$
near 1290 \AA,
to bring it to the same normalization as STIS, as described in 
\S\ref{model}.

The spectra
in Fig.~\ref{cosstisfit}
show that the shock front near $7000 \kms$ has indeed
moved during the interval between the STIS and COS observations.
One way to measure the velocity change
would be to shift the two spectra horizontally,
but the plethora of stellar lines precludes a precise measurement by this naive approach.

We follow instead a more rigorous approach
that models the physics of the situation
and that is unbiased by the presence of stellar lines.
Fig.~\ref{sicos}
shows that, at the shock front,
unshocked gas is attenuating the stellar flux by a factor of ~ 0.4.
If this is correct,
and the shock front has shifted by $\Delta ( r_s / t )$
from STIS to COS,
then the STIS spectrum should equal the COS spectrum
multiplied by an inverted top hat $1 - 0.4 = 0.6$ deep
and $\Delta ( r_s / t )$ wide.
We can think of the top hat as representing material that was absorbing in 1999 that has been ``eaten away" by the reverse shock since.
Observationally, this absorbing top hat
will be convolved with the overall LSF of the spectrum.

Fig.~\ref{cosstisfit}
shows the best-fit COS spectrum adjusted by this procedure,
with the convolved, absorbing top hat shown as a dotted line.
The fit was obtained by minimizing the $\chi^2$
between the STIS and adjusted COS spectra,
the noise being taken to be photon shot noise.
The correlations between spectral bins introduced
by convolving the spectra were taken into account.

The best fitting velocity shift between STIS (1999) and COS (2010) spectra is
\begin{eqnarray}
\label{measuredDelta}
  \Delta ( r_s / t ) & = & - 44 \pm 11 \kms, \\
  	\: {\rm or} & &\, - 4.2 \pm 1.0 \kmsy \nonumber
  \quad
  (\mbox{measured}).
\end{eqnarray}
The measured $1\, \sigma$ uncertainty of $11 \kms$
is about 60\% of a COS resolution element.    There can be little doubt that the velocity shift is real, as the measured value differs from zero by  $4 \, \sigma$.
The measured velocity shift~(\ref{measuredDelta})
is reasonably consistent with the predicted  shift~(\ref{predictedDelta}),
differing  by $16 \pm 11 \kms$, or $1.5\, \sigma$.   The value of $6982 \kms$ for $r_s/t$ given in Table~\ref{par} is just the STIS value (which is more precisely measured) reduced by $44 \kms$.

It should be remarked that we have assumed in this paper,
as previously in \citet{hamilton97,hamilton07},
that the shock front is perpendicular to the line of sight,
a plausible assumption given that the SM star is close to
the projected center of the SN~1006 remnant.
If the shock front normal were inclined by an angle $\theta$
to the line of sight, but the ejecta are moving directly away from us, then the main effect would be to increase
the observed rate of change of the free-expansion velocity by a factor $1/\cos \theta$.
If a future observation should measure a change faster than that predicted by  equation~(\ref{predictedDelta}) by a statistically significant  margin,
then one natural explanation would be that the shock front
is not  perpendicular to the line of sight.


\notetoeditor{We suggest that Fig. 5 be a full-page width (2 column) figure about here.  We would like for Fig. 4 (single-column format) and Fig. 5 both to be in color in the printed journal; if both could go on the same page, that would reduce cost.}

\section{Apparent Decrease in Shocked \SIii\ Absorption}
\label{absorption}

An unanticipated change between the two spectra
is that the optical depth in shocked \SIii\
appears to have decreased over the 10.5 year interval between STIS and COS observations.   The change   is illustrated in
Fig.~\ref{cosstisfitwide}, which compares the COS (2010) and STIS (1999) spectra
over the full range of the broad, redshifted \SIii\ absorption feature.
The ratio of COS to STIS fluxes
indicates  a small but significant flux excess, with a profile
similar to that of the shocked \SIii\ illustrated in Fig.~\protect\ref{sicos}.
As discussed in \S\ref{model},
the COS flux has first  been multiplied by a linear function
varying from
$1.03$ at $1190 \, {\rm \AA}$ to
$1.08$ at $1290 \, {\rm \AA}$
in order to bring its {\it overall\,} flux to the same level as STIS\@.

To measure the apparent change in absorption, we first adjusted the COS spectrum for the $44 \kms$ shift in the shock front as described in \S\ref{evolution}, so that the two should be comparable.   We then assumed  that the COS:STIS ratio has the same profile as shocked \SIii\ but with unknown amplitude, and did a one-parameter least-squares fit to determine that  amplitude.
This procedure is more precise than simply taking the ratio of
shocked column densities in Table~\ref{par},
whose values are model-dependent.
We find that 
the column density of shocked \SIii\
measured from the COS spectrum is
$0.93 \pm 0.02$
times that in the STIS spectrum from ten years earlier.    The statistically significant $7 \pm 2\%$  change
is larger than the $\sim 1\%$ change one might simplistically have expected
over 10~years of observation of a 1000~year old supernova remnant.    The 2\% uncertainty includes only statistical uncertainty, and not any systematic uncertainty associated with our modeling of the continuum.  The 7\% change is comparable to the $\sim 8\%$ systematic adjustment we have made to the COS flux, and  some systematic  uncertainty attends our modeling of the relative sensitivities between COS and STIS.

If the unexpectedly rapid decrease in optical depth of shocked \SIii\ is real,
then a possible explanation is that the shock has been ploughing through a
clump of Si.
Numerical simulations of shocks overrunning interstellar clouds \citep{klein94}
indicate that clouds of sufficiently high density contrast
can undergo significant pressure and density transients
as the shock propagates through the cloud.
The line profile of {\it unshocked} \SIii\ illustrated in Figure~\ref{sicos}
does indicate a density decreasing inward 
by a factor of 2 over an interval of about $700 \kms$.
This change is equivalent to about 2\% in $30 \kms$,
which is several times smaller than required by the observed change
in optical depth.
The profile of unshocked  \SIii\ shown in Figure~\ref{sicos}
does show possible smaller scale structure consistent with more rapid variation,
but the reality of such structure is confused by the many stellar lines.

\section{Summary}
Using UV absorption-line spectra against the SM star taken with STIS in 1999 and with COS in 2010, we have probed changes in the absorbing supernova ejecta seen along a ``core sample" through the SN~1006 remnant.   At both epochs we see strong absorption due to \SIii, all of it red-shifted and with a sharp red edge.    Following \citet{hamilton97, hamilton07}, we have interpreted  the profile of the strong 1260 \AA\ absorption feature as resulting from both  unshocked  and shocked  (by the reverse shock) \SIii\ ejecta, with the sharp red edge to this feature being due to freely expanding ejecta just before being hit by the reverse shock.  After carefully taking into account small systematic differences in both wavelength and flux between measurements with the two instruments, we find that the red absorption edge has shifted by $-44 \pm 11\kms$ over the 10.5 yr interval between observations, or $-4.2 \pm 1.0 \kmsy $.   This measurement confirms the long-held expectation that in a young SNR,  the reverse shock will encounter ever-slower material as it works its way inward through the supernova ejecta.   We also find the unanticipated result that the optical depth of shocked \SIii\ has decreased by $7 \pm 2 \%$, faster than might be expected to result from homologous expansion over 10 years in the SNR's 1000-yr history.    Clumped Si ejecta provides a possible explanation both for the decrease in optical depth and for the absence of corresponding blue-shifted \SIii\ absorption.

\acknowledgments
We appreciate the careful reading and thoughtful suggestions by the (anonymous) referee, which have helped clarify this paper. 
Support for this research has been provided  by NASA Grant HST-GO-11659 from the Space Telescope Science Institute, which is operated by AURA, Inc. under NASA contract 5-26555\@.   P.F.W. acknowledges additional support from NSF grant AST-0908566.

\bibliographystyle{apj}







\begin{deluxetable}{ccccrr}
\tablecaption{Observations of the Schweizer-Middleditch star from {\it HST} }
\tablehead{\colhead{Instrument} & 
 \colhead{Obs.~Date} & 
 \colhead{Grating} & 
 \colhead{FWHM} &
 \colhead{Exposure} &
 \colhead{{\it HST} Program}
\\
\colhead{~} & 
 \colhead{~} & 
 \colhead{~} & 
 \colhead{($\kms$)} & 
 \colhead{(s)} &
\colhead{ID}  
}
\scriptsize
\tablewidth{0pt}\startdata
STIS &  1999 Jul 24-25 &  E140M &  4.6 & 23,186\phn\phn &  7349\phn\phn \\ 
COS &  2010 Jan 16 &  G130M &  18. & 1,756\phn\phn &  11659\phn\phn \\ 
\hline
\enddata 
\label{tab_obs}
\end{deluxetable}

 \begin{deluxetable}{lcc}
 \tablecaption{Measured parameters}
\tablehead{\colhead{Parameter} & 
\colhead{COS (this paper)} & 
\colhead{STIS (HFB07)}
}
\tablewidth{0pt}\startdata
\label{par}
$r_s/t$ & $6982 \pm 11 \kms$ & $7026 \pm 3 \, (\mbox{rel.})\pm 10 \, (\mbox{abs.}) \kms$
\\
$\overline v$
 & $5030 \pm 70 \kms$
 & $5160 \pm 70 \kms$
\\
$\sigma$
 & $1140 \pm 50 \kms$
 & $1160 \pm 50 \kms$
\\
$v_s$
 & $2630 \pm 120 \kms$
 & $2680 \pm 120 \kms$
\\
$N^{\rm shk}_{\rm Si II}$
 & $7.4 \pm 0.3 \times 10^{14} \, {\rm cm}^{-2}$
 & $8.6 \pm 0.3 \times 10^{14} \, {\rm cm}^{-2}$
\\
\hline
\enddata
    \end{deluxetable}

\begin{figure}
\plotone{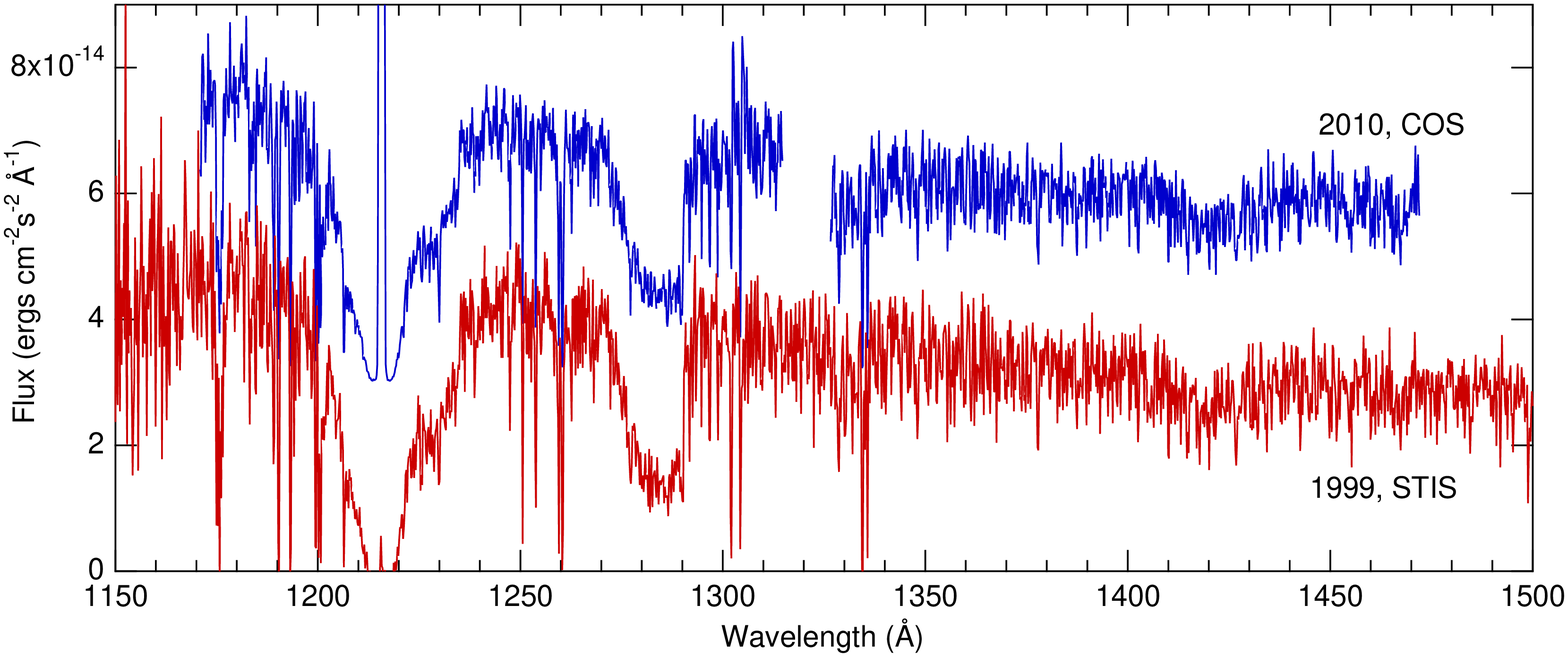}
\caption{
The complete COS and STIS FUV spectra of the SM star.  
The absorption feature at 1217 \AA\  has 
contributions from geocoronal Ly$\alpha$\  emission, stellar and interstellar Ly$\alpha$\   absorption, redshifted \SIii\ 1190.4 \AA\  and 1193.3 \AA\ absorption, and   redshifted \SIiii\ 1206.5 \AA\ absorption. The most prominent ``clean" feature is the 
broad, redshifted \SIii\ 1260.4 \AA\ line, due to a combination of shocked and unshocked  \SIii\ 
ejecta.  The COS spectrum has been displaced vertically for clarity. }
\end{figure}
\newpage

\begin{figure}
\epsscale{.6}
\plotone{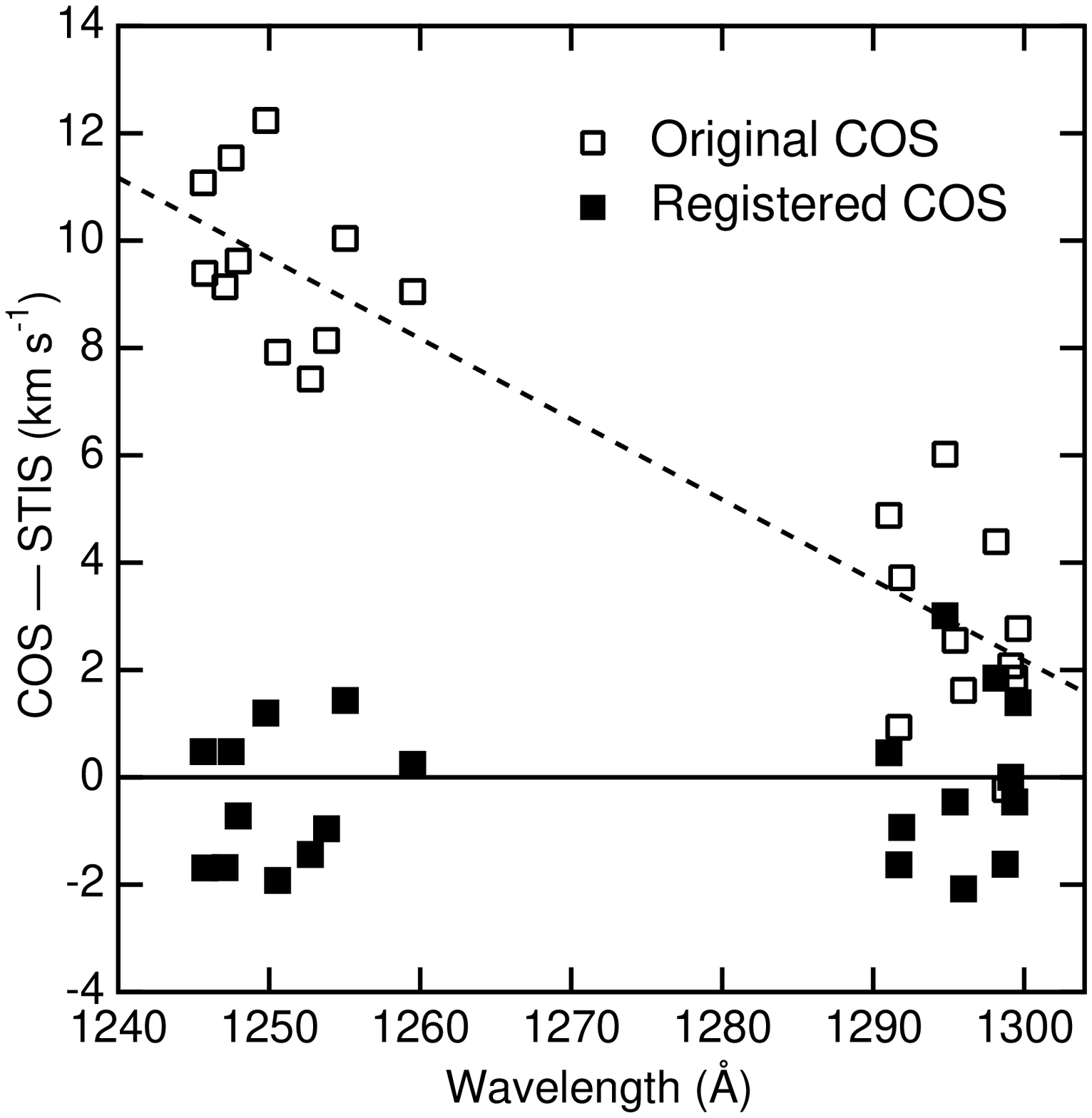}
\caption{
The difference between  measured wavelengths of several narrow lines, on both blue and red sides of the \SIii\ absorption feature, is plotted as a function of wavelength.  A simple linear transformation in the COS wavelength scale  brings the two spectra into registration, with an RMS uncertainty of $\pm 1.5\kms$.
 }
\end{figure}
\pagebreak

\begin{figure}
    \begin{center}
    \includegraphics[scale=0.8]{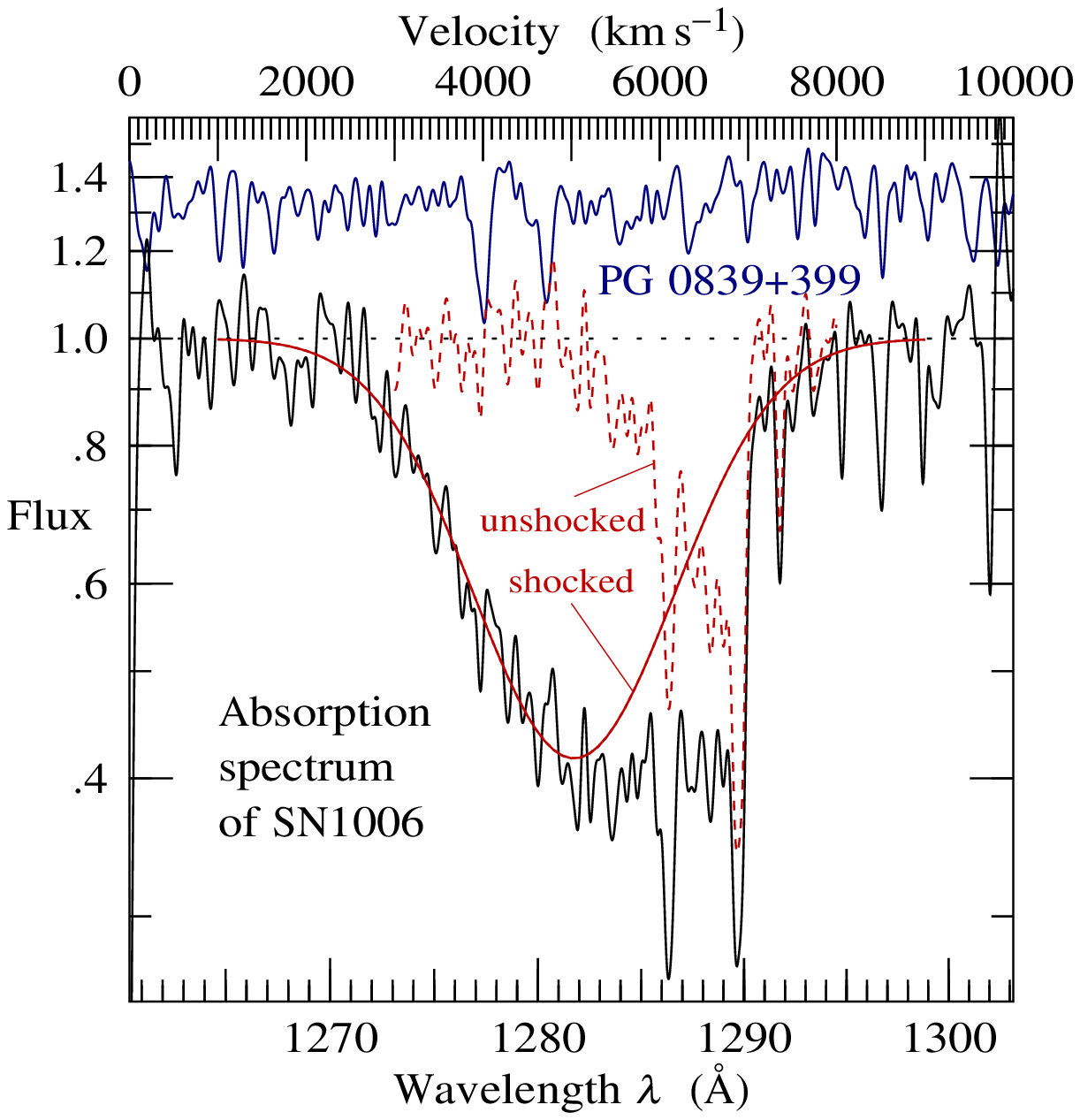}
    \end{center}
    \caption[1]{
    \label{sicos}
COS absorption spectrum of SN~1006
around the redshifted \SIii\ $1260 \, {\rm\AA}$ feature,
showing the best-fit Gaussian profile of shocked \SIii,
and the residual unshocked \SIii.
The spectrum is the ratio of
the interstellar-line-excised,
continuum-corrected
COS spectrum of the SM star
to the STIS spectrum of the comparision star PG\ 0839+399, with
each smoothed to a resolution of $80 \kms$ FWHM
before their ratio was taken.
The upper spectrum shows the
interstellar-line-excised,
continuum-corrected comparison stellar spectrum
of PG\ 0839+399 at the same resolution,
shifted by
$+ 32 \kms$ to mesh its stellar lines with those of the SM star,
and offset vertically to separate it from the SN~1006 spectrum.
    }
    \end{figure}

    \begin{figure}
    \begin{center}
    \includegraphics[scale=0.8]{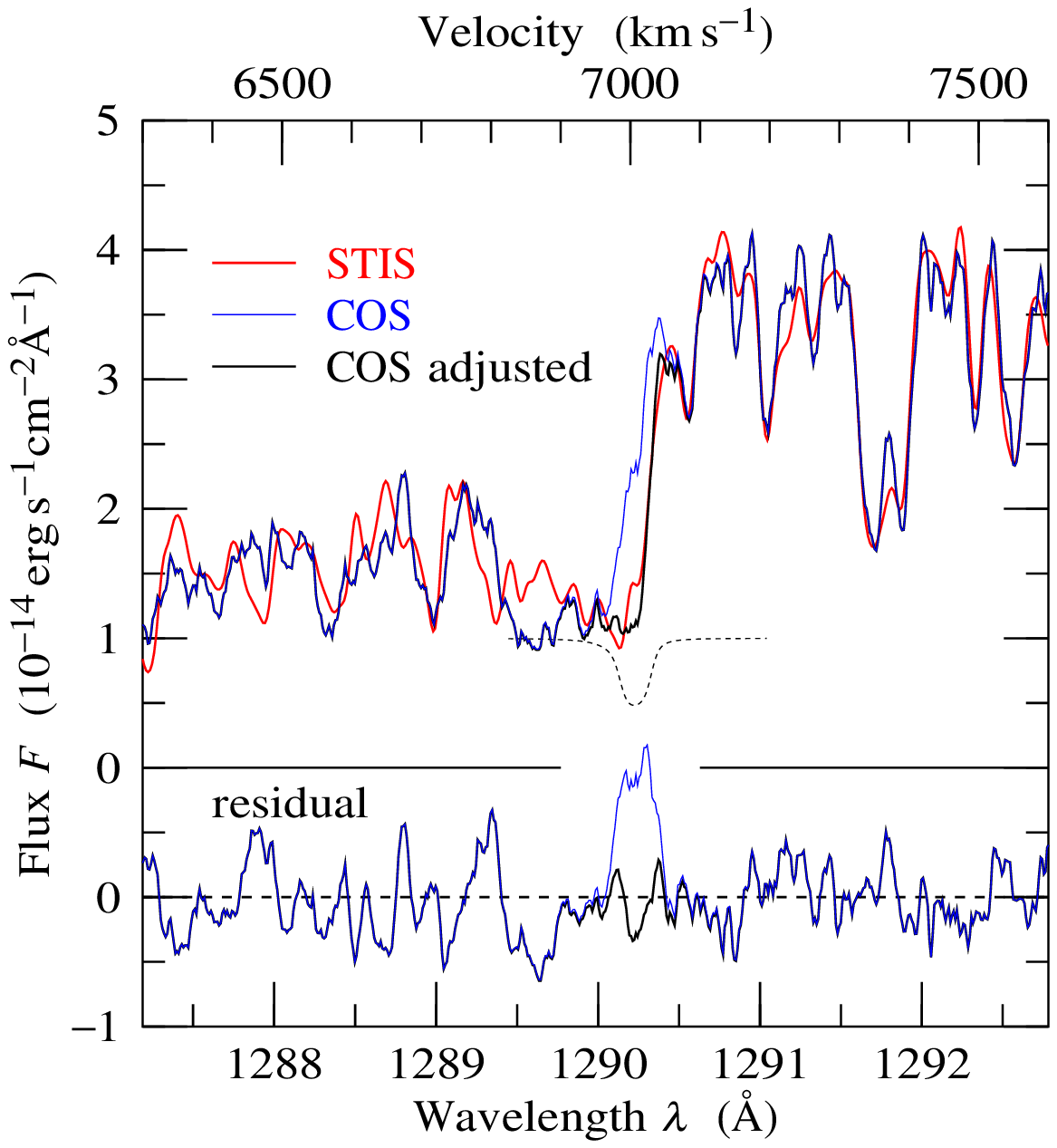}
    \end{center}
    \caption[1]{
    \label{cosstisfit}
Comparison of COS (blue) and STIS (red) spectra of the SM star behind SN~1006.
To permit direct comparison,
each spectrum has been convolved with the line spread functon of the other.
Also shown is the best-fit adjusted COS spectrum (black),
in which the COS flux has been attenuated by a factor of $0.4$
over a slice of velocity $44 \kms$ wide starting
at the $7026 \kms$ velocity of the shock front observed with STIS\@.
The dotted line shows the factor by which the COS spectrum
has been multiplied,
an inverted top hat convolved with the overall LSF.
The residual difference between the COS and STIS spectra is shown.
    }
    \end{figure}

    \begin{figure*}[t!]
    \begin{minipage}{7in}
    \begin{center}
    \leavevmode
    \includegraphics[scale=.94]{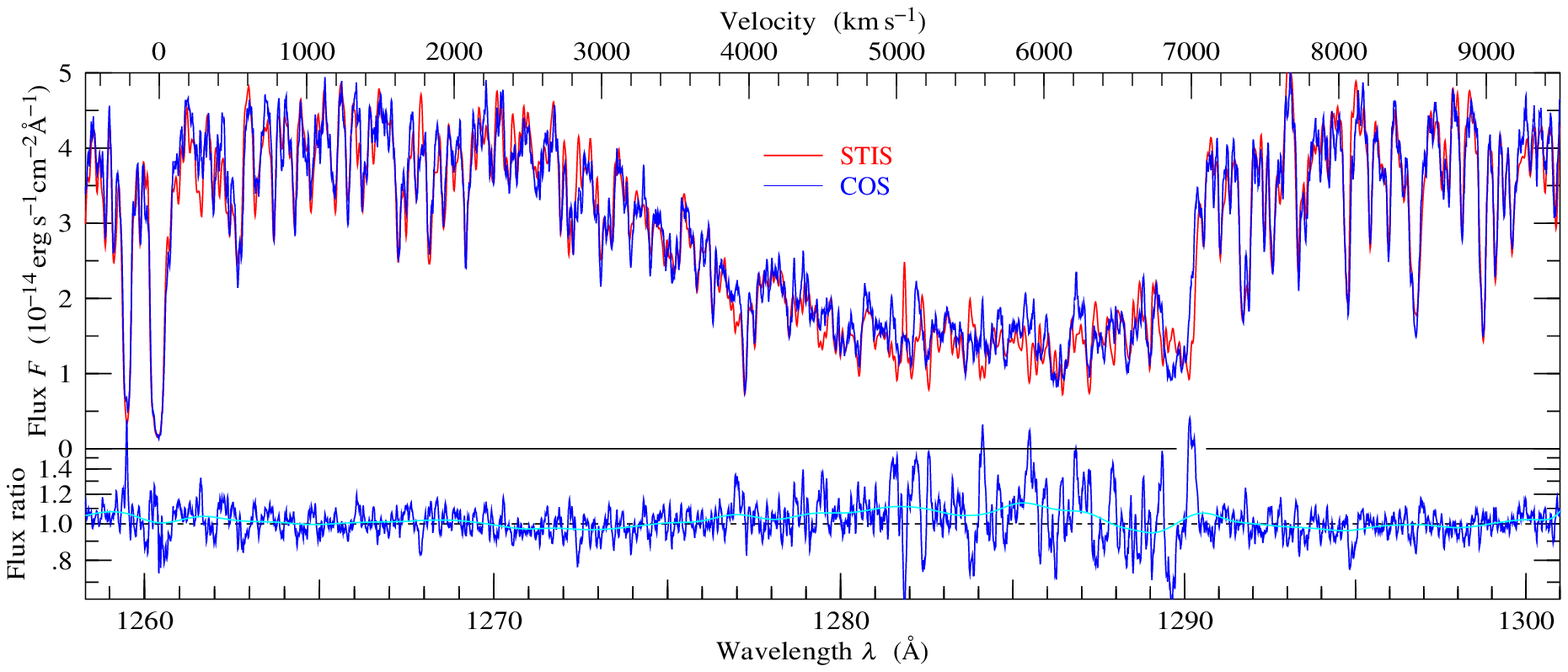}
    \end{center}
    \caption[1]{
    \label{cosstisfitwide}
Comparison of COS (blue) and STIS (red) spectra of the SM star behind SN1006.
Each spectrum has been convolved with the line spread function of the other.
The lower panel shows the ratio of the COS to STIS fluxes.
The smooth (cyan) line is the ratio smoothed with a gaussian to
$320 \kms$ FWHM.
The ratio shows a small but significant excess with a profile
similar to that of the shocked \SIii\ illustrated in Figure~\protect\ref{sicos},
indicating a decrease in the optical depth of shocked \SIii\
in COS compared to STIS\@.
    }
    \end{minipage}
    \end{figure*}

\end{document}